\documentclass[9pt,twocolumn,twoside]{optica}
\usepackage[normalem]{ulem}
\usepackage{amsmath}
\usepackage{lipsum}
\newcommand{\stkout}[1]{\ifmmode\text{\sout{\ensuremath{#1}}}\else\sout{#1}\fi}
\renewcommand{\phi}{\varphi}
\setboolean{shortarticle}{false}
\setboolean{minireview}{false}

\usepackage{color}

\title{Angular momenta, helicity, and other properties of dielectric-fiber and metallic-wire modes}

\author[1,*]{M. F. Picardi}
\author[2,3,*]{K. Y. Bliokh}
\author[1]{F. J. Rodríguez-Fortuño}
\author[4]{F.~Alpeggiani}
\author[2,5]{F. Nori}
\affil[1]{Department of Physics, King’s College London, London WC2R 2LC, UK}
\affil[2]{Theoretical Quantum Physics Laboratory, RIKEN Cluster for Pioneering Research, Wako-shi, Saitama 351-0198, Japan}
\affil[3]{Nonlinear Physics Centre, RSPE, The Australian National University, Canberra, ACT 0200, Australia}
\affil[4]{Kavli Institute of Nanoscience, Delft University of Technology, 2600 GA Delft, The Netherlands}
\affil[5]{Physics Department, University of Michigan, Ann Arbor, Michigan 48109-1040, USA}

\affil[*]{These authors contributed equally to this work}

\ociscodes{(260.0260) Physical optics; (260.6042) Singular optics; (240.6680) Surface plasmons; (060.2310) Fiber optics}

\begin{abstract}
Spin and orbital angular momenta (AM) of light are well studied for \emph{free-space} electromagnetic fields, even nonparaxial. One of the important applications of these concepts is the information transfer using AM modes, often via optical fibers and other guiding systems. However, the self-consistent description of the spin and orbital AM of light in \emph{optical media} (including dispersive and metallic cases) was provided only recently [K.Y. Bliokh \textit{et al.}, \textit{Phys. Rev. Lett.} \textbf{119}, 073901 (2017)]. Here we present the first accurate calculations, both analytical and numerical, of the spin and orbital AM, as well as the helicity and other properties, for the full-vector eigenmodes of cylindrical dielectric and metallic (nanowire) waveguides. We find remarkable fundamental relations, such as the \emph{quantization of the canonical total AM} of cylindrical guided modes in the general nonparaxial case. This quantization, as well as the noninteger values of the spin and orbital AM, are determined by the generalized \emph{geometric and dynamical phases} in the mode fields. Moreover, we show that the spin AM of metallic-wire modes is determined, in the geometrical-optics approximation, by the \emph{transverse spin} of surface plasmon-polaritons propagating along helical trajectories on the wire surface. Our work provides a solid platform for future studies and applications of the AM and helicity properties of guided optical and plasmonic waves.
\end{abstract}

\setboolean{displaycopyright}{false}
\begin{document}

\maketitle

%%%%%%%%%%%%%%%%%%%%%%%%%%%%%%%%%%%%%%%%%%
\section{Introduction}
%%%%%%%%%%%%%%%%%%%%%%%%%%%%%%%%%%%%%%%%%%
%
Spin and orbital angular momentum (AM) of light are well-established concepts in modern optics (see, e.g., books \cite{allen2003optical,torres2011twisted,andrews2012angular} and reviews \cite{allen1999iv,yao2011orbital,bliokh2015transverse}). Despite some subtle issues originating from quantum and field-theory aspects \cite{berestetskii1982quantum,soper2008classical,leader2016photon}, the spin and orbital AM, as well as their local densities, are well-defined for monochromatic electromagnetic fields (even nonparaxial) \emph{in free space} \cite{bliokh2015transverse,van1994commutation,bliokh2010angular,barnett2010rotation,bialynicki2011canonical,bliokh2014conservation}. In parallel with theoretical studies, the spin and orbital AM were intensively explored experimentally. In the past decades, these have found numerous applications in diverse areas including optical manipulations \cite{he1995direct,gahagan1996optical,garces2003observation,grier2003revolution}, quantum optics \cite{mair2001entanglement,leach2002measuring,leach2010quantum}, information transfer and communications \cite{gibson2004free,wang2012terabit,tamburini2012encoding}.

Importantly, vortex modes carrying AM naturally appear in cylindrically-symmetric waveguides, such as dielectric fibers \cite{snyder2012optical,marcuse1972light} or metallic wires \cite{pfeiffer1974surface,novotny1994light}. Moreover, one of the important applications of the optical AM is the multi-channel information transfer via optical fibers \cite{bozinovic2013terabit,willner2015optical}. However, the rigorous characterization of the spin and orbital AM of a multimode waveguide still remains an unsolved problem involving nonparaxial electromagnetic fields in \emph{inhomogeneous media}. It is known that fiber modes exhibit various spin-orbit interaction phenomena \cite{dooghin1992optical,liberman1992spin,bliokh2004modified,alexeyev2007fiber, golowich2014asymptotic}, i.e., coupling between the polarization and orbital degrees of freedom \cite{bliokh2015spin}. Furthermore, the total AM must be conserved due to the cylindrical symmetry of the system \cite{alexeyev1998optical,gregg2015conservation}. However, none of these studies answers the question “\emph{what are the spin and orbital AM values?}” for the cylindrical guided modes.

The only work that properly addressed the above question \cite{le2006angular} did this for the simplest situation of a single fundamental mode of a nondispersive (dielectric) nanofiber. Moreover, only the electric-field (but not the magnetic-field) contributions to the energy, spin, and orbital AM of the fiber mode were considered there. Importantly, Ref. \cite{le2006angular} demonstrated that the problem of the characterization of the AM of the guided modes is closely related to the \emph{Abraham–Minkowski dilemma} in the characterization of the momentum of light in a medium \cite{brevik1979experiments,pfeifer2007colloquium,barnett2010enigma,milonni2010momentum,kemp2011resolution}. Traditionally, this dilemma discussed only the linear momentum of plane waves in homogeneous media, and only very recently it was solved for the momentum, spin, and orbital AM of arbitrary monochromatic fields in inhomogeneous and dispersive (but isotropic and lossless) media \cite{bliokh2017optical,bliokh2017optical2}. In particular, it was shown that the canonical (Minkowski-type) momentum, spin and orbital AM acquire very natural forms similar to the well-known Brillouin energy density \cite{jackson1999classical,landau2013electrodynamics}.

In this work, we show that the general description \cite{bliokh2017optical,bliokh2017optical2} of the momentum and AM of light works perfectly for cylindrical modes in both dielectric and metallic (plasmonics) waveguides. This allows one to unambiguously quantify all dynamical properties of complex eigenmodes in inhomogeneous dispersive structures. In particular, we find a very simple yet fundamental result: the canonical total (spin + orbital) AM of the eigenmodes of cylindrical waveguides always takes on \emph{integer values} $\ell$ (the topological charge of the vortex in the longitudinal field components) in units of $\hbar$ per photon. Note that this simple result cannot be obtained within the usual Poynting-vector-based (i.e., kinetic or Abraham) formalism \cite{jackson1999classical,landau2013electrodynamics}, where the total AM is non-integer. Thus, our approach allows one to extend the results and intuition developed for free-space fields (where the total AM of cylindrical modes is integer \cite{van1994commutation,bliokh2010angular,bliokh2014conservation}) to the fields in inhomogeneous dispersive media. Remarkably, we show that, akin to earlier free-space results \cite{bliokh2010angular}, the non-integer spin and orbital AM values for guided modes is closely related to the generalized \emph{geometric phases} in the mode fields. Moreover, for metallic-wire modes we provide a simple geometrical-optics model based on the helical rays of surface plasmon-polaritons. It shows that the longitudinal spin AM of the metallic-wire modes originates from the transverse spin \cite{bliokh2015transverse,bliokh2017optical,bliokh2017optical2,bliokh2012transverse,aiello2015transverse} of skew surface plasmon-polaritons.

We also show that the canonical \cite{bliokh2017optical,bliokh2017optical2} and kinetic (Poynting-Abraham) \cite{jackson1999classical,landau2013electrodynamics} momentum of the guided modes can be associated with the propagation constant $\beta$ and the group velocity $\partial\omega/\partial\beta$, respectively. Last but not least, we also examine the \emph{helicity} of guided modes. This is an independent fundamental quantity (conserved in free space), which is equivalent to the spin AM only in the simplest plane-wave case, but generally it characterizes \emph{the degree of chirality} of the electromagnetic field \cite{afanasiev1996helicity,trueba1996electromagnetic,cameron2012optical,bliokh2013dual,fernandez2013electromagnetic}. Akin to the AM, the description of the optical helicity was extended from free space to dispersive inhomogeneous media only very recently \cite{van2016conditions,alpeggiani2018electromagnetic}. We show that the helicity of guided modes differs from their spin AM and can take any values in the $(-1,1)$ range (in units of $\hbar$ per photon). This shows that the cylindrical guided modes are the eigenmodes of the longitudinal component of the total AM (with integer eigenvalues), but not helicity eigenstates.

We perform both analytical and numerical calculations for dielectric multimode fibers, as well as for metallic wires supporting plasmonic modes. Our results reveal fundamental features of the momentum, AM, and helicity properties, universal for electromagnetic modes in various complex media.

%%%%%%%%%%%%%%%%%%%%%%%%%%%%%%%%%%%%%%%%%%
\section{Basic equations and guided-modes properties}
\label{sec:basic}
%%%%%%%%%%%%%%%%%%%%%%%%%%%%%%%%%%%%%%%%%%
\subsection{Energy, momentum, angular momentum, and helicity}\label{subsec:definitions}
%%%%%%%%%%%%%%%%%%%%%%%%%%%%%%%%%%%%%%%%%%
%
Recently, an efficient formalism describing canonical dynamical properties (momentum, angular momentum, etc.) of monochromatic electromagnetic fields in isotropic dispersive media was developed \cite{bliokh2017optical,bliokh2017optical2}. According to this, the cycle-averaged energy (Brillouin expression \cite{jackson1999classical,landau2013electrodynamics}), momentum, spin, orbital, and total AM densities in the field can be written as:
\begin{eqnarray}
\label{eq:definitions}
&&W=\frac{1}{4}\!\left(\tilde{\varepsilon}|\mathbf{E}|^2+\tilde{\mu}|\mathbf{H}|^2\right),~~ 
\nonumber \\
&&\mathbf{P}=\frac{1}{4\omega}\mathrm{Im}\!\left[\tilde{\varepsilon}\mathbf{E}^*\!\cdot(\nabla)\mathbf{E}+\tilde{\mu}\mathbf{H}^*\!\cdot(\nabla)\mathbf{H}\right], \\
&&\mathbf{S}=\frac{1}{4\omega} \mathrm{Im}\!\left( \tilde{\varepsilon}\mathbf{E}^*\!\times\mathbf{E} +\tilde{\mu}\mathbf{H}^*\!\times\mathbf{H} \right),~~~ 
\mathbf{L}=\mathbf{r}\times\mathbf{P}, ~~~ 
\mathbf{J}=\mathbf{L}+\mathbf{S}. \nonumber
\end{eqnarray}
Here, $\mathbf{E(r)}$ and $\mathbf{H(r)}$ are the complex electric and magnetic field amplitudes, $\omega$ is the frequency, and $(\tilde{\varepsilon},\tilde{\mu})=(\varepsilon,\mu)+\omega\, d(\varepsilon,\mu)/d\omega$ are the dispersion-modified permittivity $\varepsilon$ and permeability $\mu$ of the medium, which are assumed to be real. In Eqs. \ref{eq:definitions} and in what follows we neglect inessential common factors and use the dimensionless parameters $(\varepsilon,\mu)$ in Gaussian units [which should be understood as $(\varepsilon,\mu)\rightarrow(\varepsilon_0\varepsilon,\mu_0\mu)$ in SI units].

The quantities (\ref{eq:definitions}) represent \emph{canonical Minkowski-type} properties of the field \cite{bliokh2017optical2,bliokh2017optical}. In particular, the canonical momentum density $\mathbf{P}$ can naturally be associated with the local wavevector (phase gradient) in the field: $\mathbf{P}/W=\mathbf{k}_{\rm loc}/\omega$. In turn, the \emph{kinetic Abraham} momentum density is given by the Poynting vector \cite{jackson1999classical,landau2013electrodynamics}:
\begin{equation}
\label{eq:poynting}
\boldsymbol{\mathcal{P}}=\frac{1}{2c}\mathrm{Re}\left(\mathbf{E}^*\!\times\mathbf{H}\right).
\end{equation}
($c\rightarrow 1$ in SI units). The Poynting-Abraham momentum density actually describes the energy flux and the \emph{group velocity} of the wave propagation. For localized modes with well-defined real wave vector (phase gradient) $\mathbf{k}$, the group velocity is given by the ratio of the integral Poynting vector and Brillouin energy \cite{snyder2012optical,marcuse1972light,bliokh2017optical,bliokh2017optical2,jackson1999classical}: $\mathbf{v}_g=\partial\omega/\partial\mathbf{k}=c^2\left\langle \boldsymbol{\mathcal{P}}\right\rangle/\left\langle W\right\rangle$, where $\langle ... \rangle$ denotes the integration over the corresponding coordinates. Note that the Poynting vector (\ref{eq:poynting}) also determines the \emph{kinetic} (Abraham-type) total AM density \cite{bliokh2017optical2,jackson1999classical}: 
\begin{equation}
\label{eq:kinetic_am_desity}
\boldsymbol{\mathcal{J}}=\mathbf{r}\times\boldsymbol{\mathcal{P}}.
\end{equation}
As we show below, for the waveguide modes its properties differ considerably from the canonical AM (\ref{eq:definitions}). In particular, even their integral values differ, $\left\langle \boldsymbol{\mathcal{J}}\right\rangle \neq \left\langle \boldsymbol{J}\right\rangle$, in contrast to the free-space situation \cite{fernandez2013electromagnetic}. The physical difference between the kinetic-Abraham and canonical-Minkowski quantities is that the former ones describe the properties of \emph{electromagnetic fields only}, while the latter ones characterize properties of the \emph{whole wave mode} (i.e., a polariton, which involves, on the microscopic level, oscillations of both fields and electrons in matter) \cite{bliokh2017optical2,partanen2017photon}. In fact, the concept of “photon in a medium” implies such polariton excitation characterized by Minkowski-type quantities. Moreover, it is the canonical-Minkowski quantities that are \emph{conserved} in media with the corresponding symmetries \cite{bliokh2017optical,bliokh2017optical2,philbin2011electromagnetic,philbin2012optical}.

The electromagnetic \emph{helicity} is an independent important property of electromagnetic fields, which is related to the “dual symmetry” between the electric and magnetic fields \cite{afanasiev1996helicity,trueba1996electromagnetic,cameron2012optical,bliokh2013dual,fernandez2013electromagnetic,van2016conditions}. It quantifies the chirality of the field, and generally differs from the spin AM. Recently, it was shown \cite{alpeggiani2018electromagnetic} that the helicity density in dispersive inhomogeneous dielectrics and metals can be written as:
\begin{equation}
\label{eq:helicity}
\mathfrak{S}=\frac{1}{2\omega}|\tilde{n}|\,
\mathrm{Im}\!\left(\mathbf{H}^*\!\cdot\mathbf{E}\right) 
= \frac{1}{4\omega}\left|\tilde{\varepsilon}\sqrt{\frac{\mu}{\varepsilon}} +\tilde{\mu}\sqrt{\frac{\varepsilon}{\mu}}\right|
\mathrm{Im}\!\left(\mathbf{H}^*\!\cdot\mathbf{E}\right),
\end{equation}
where $\tilde{n}=\sqrt{\varepsilon\mu}+d\sqrt{\varepsilon\mu}/d\omega$ is the group refractive index of the medium. For dispersionless dielectrics, $\tilde{n}=\sqrt{\varepsilon\mu}$, while for Drude-model metals with $\varepsilon=1-\omega_p^2/\omega^2$, $\tilde{\varepsilon}=2-\varepsilon$ ($\omega_p$ is the plasma frequency), and $\tilde{\mu}=\mu$, one has $|\tilde{n}|=\sqrt{|\mu/\varepsilon|}$.

Below we investigate the momentum, AM, and helicity properties of the eigenmodes of cylindrical dielectric fibers and metallic wires. We will calculate the normalized values “per photon in units of $\hbar$ ”, which are given by the local density ratios $\omega\mathbf{S}/W$, $\omega\mathbf{L}/W$, $\omega\mathfrak{S}/W$, etc., and by the corresponding integral ratios $\omega\left\langle\mathbf{S}\right\rangle/\left\langle W \right\rangle$, etc.

%%%%%%%%%%%%%%%%%%%%%%%%%%%%%%%%%%%%%%%%%%
\subsection{Eigenmodes of cylindrical fibers and wires}
\label{subsec:eigenmodes}
%%%%%%%%%%%%%%%%%%%%%%%%%%%%%%%%%%%%%%%%%%
%
We consider a cylindrical non-magnetic medium of radius $r_0$ in vacuum, which is characterized by the permittivity and permeability:
\begin{equation*}
  \varepsilon=\left\{
  \begin{array}{@{}ll@{}}
    \varepsilon_1, & \text{for}~~~ r<r_0\\
    \varepsilon_2, & \text{for}~~~ r>r_0
  \end{array}\right. \quad \text{and} \quad \mu=1.
\end{equation*} 
(We, however, keep $\mu$ in the equations to facilitate the transition to SI units: $\varepsilon\rightarrow\varepsilon_0\varepsilon$, $\mu\rightarrow\mu_0$.) In dielectric waveguides the dispersion is neglected, so that $\tilde{\varepsilon}=\varepsilon$ and $\varepsilon_1>\varepsilon_2$, while in metallic wires $-\varepsilon_1>\varepsilon_2>0$, but $\tilde{\varepsilon}_1>\varepsilon_2>0$. In what follows, we assume the Drude plasma dispersion for the metal: $\varepsilon_1(\omega)=1-\omega_p^2/\omega^2$.

\begin{figure}[t]
\centering
\fbox{\includegraphics[width=0.9\linewidth]{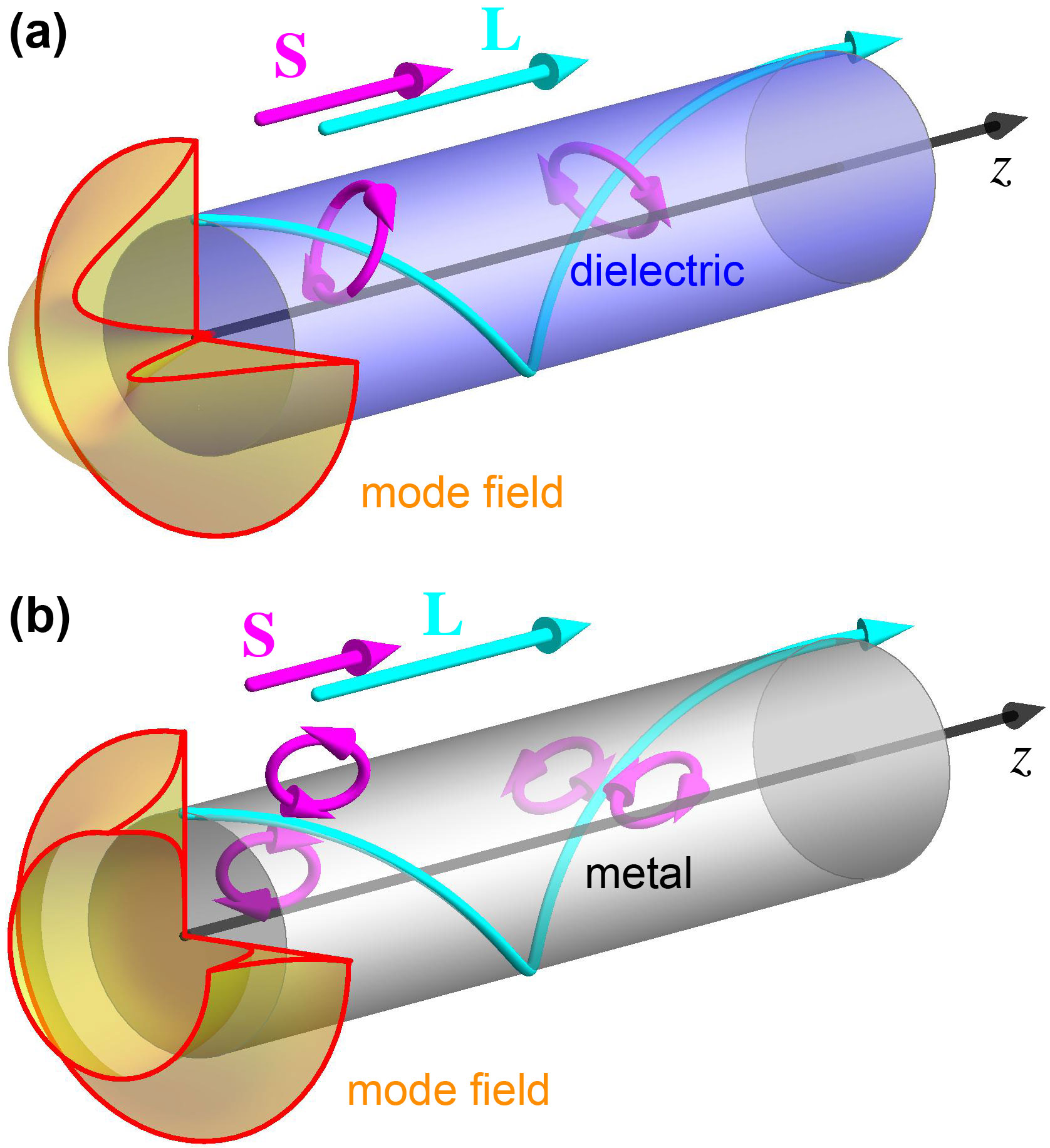}}
\caption{Schematic pictures of the eigenmodes of a dielectric fiber (a) and a metallic
wire (b). The geometrical-optics skew rays with their polarizations (transverse
circular in dielectrics and in-plane elliptical for surface plasmon-polaritons
\cite{bliokh2015transverse,bliokh2015spin,bliokh2012transverse,aiello2015transverse}) are shown by cyan and magenta, respectively. These helical rays and their corresponding polarizations illustrate the origin of the orbital ($\mathbf{L}$) and spin ($\mathbf{S}$) AM of the cylindrical guided modes.}
\label{fig:fig0}
\end{figure}

The eigenmodes of cylindrical waveguides are well studied \cite{snyder2012optical,marcuse1972light,pfeiffer1974surface,novotny1994light}, and are schematically shown in Fig.~\ref{fig:fig0}.
Usually, the mode fields are presented using the components attached to the cylindrical coordinates $(r,\phi,z)$. However, we found that these acquire a particularly laconic form in the basis of circular polarizations attached to the transverse Cartesian coordinates: $E^\pm = (E_x\mp i E_y)/\sqrt{2}$, $H^\pm=(H_x\mp i H_y)/\sqrt{2}$. Namely, the eigenmode field inside the waveguide $(r<r_0)$ can be written as:
\begin{eqnarray}
\label{eq:eigenmode}
&&E^\pm = -\frac{i}{\sqrt{\varepsilon}\kappa}(\pm\beta A+ i k B)\,
J_{\ell \mp 1}(\rho)\,e^{i(\ell \mp 1)\phi + i \beta z}, \nonumber\\
&&H^\pm = -\frac{i}{\sqrt{\mu}\kappa}(\pm \beta B - i k A )\,
J_{\ell \mp 1}(\rho)\,e^{i(\ell \mp 1)\phi + i \beta z}, \\
&&E_z = \sqrt{\frac{2}{\varepsilon}}\,A\, J_{\ell}(\rho)\,e^{i \ell \phi + i \beta z}, \quad 
H_z = \sqrt{\frac{2}{\mu}}\,B\, J_{\ell}(\rho)\, e^{i \ell \phi + i \beta z}. \nonumber
\end{eqnarray}
Here, $k=\sqrt{\varepsilon\mu}\,\omega/c$ is the wave number in the medium, $\beta > k_0$ is the mode propagation constant ($k_0$ is the wave number in vacuum), $\kappa = \sqrt{k^2-\beta^2}$ is the radial wave number, $\rho = \kappa r$, $\ell = 0, \pm 1, \pm 2, ...$ is the azimuthal quantum number, and $J_\alpha (\rho)$ is the Bessel function of the first kind. The values of the propagation constant $\beta$ for given other parameters ($\omega$, $r_0$, etc.) are found from the transcendental characteristic equation, whereas the complex constants $A$ and $B$
are determined from the boundary conditions at $r=r_0$ (see Appendix A) \cite{marcuse1972light}. The eigenmode fields outside the fiber are given by Eqs.~(\ref{eq:eigenmode}) with the substitution: 
\begin{equation}\label{eq:eigenmode_out}
J_\alpha(\rho)\rightarrow H_\alpha^{(1)}(\rho), \quad (A,B)\rightarrow(C,D),
\end{equation}
where $H_\alpha^{(1)}(\rho)$ is the Hankel function, the radial wave number becomes imaginary, $\kappa = \sqrt{k^2-\beta^2}=i\sqrt{\beta^2-k^2}$, whereas the complex constants $C$ and $D$ are determined from the boundary conditions (see Appendix A). Equations (\ref{eq:eigenmode}) and (\ref{eq:eigenmode_out}) describe the eigenmodes of dielectric fibers \cite{snyder2012optical,marcuse1972light} and metallic wires \cite{pfeiffer1974surface,novotny1994light}. In the latter case, $\varepsilon_1 <0$, and both $k$ and $\kappa$ become imaginary inside the wire.

%%%%%%%%%%%%%%%%%%%%%%%%%%%%%%%%%%%%%%%%%%
%%%%%%%%%%%%%%%%%%%%%%%%%%%%%%%%%%%%%%%%%%

%%%%%%%%%%%%%%%%%%%%%%%%%%%%%%%%%%%%%%%%%%
\subsection{Labelling the modes with quantum numbers}
\label{subsec:labelling}
\begin{figure*}[t]
\centering
\fbox{\includegraphics[width=0.9\linewidth]{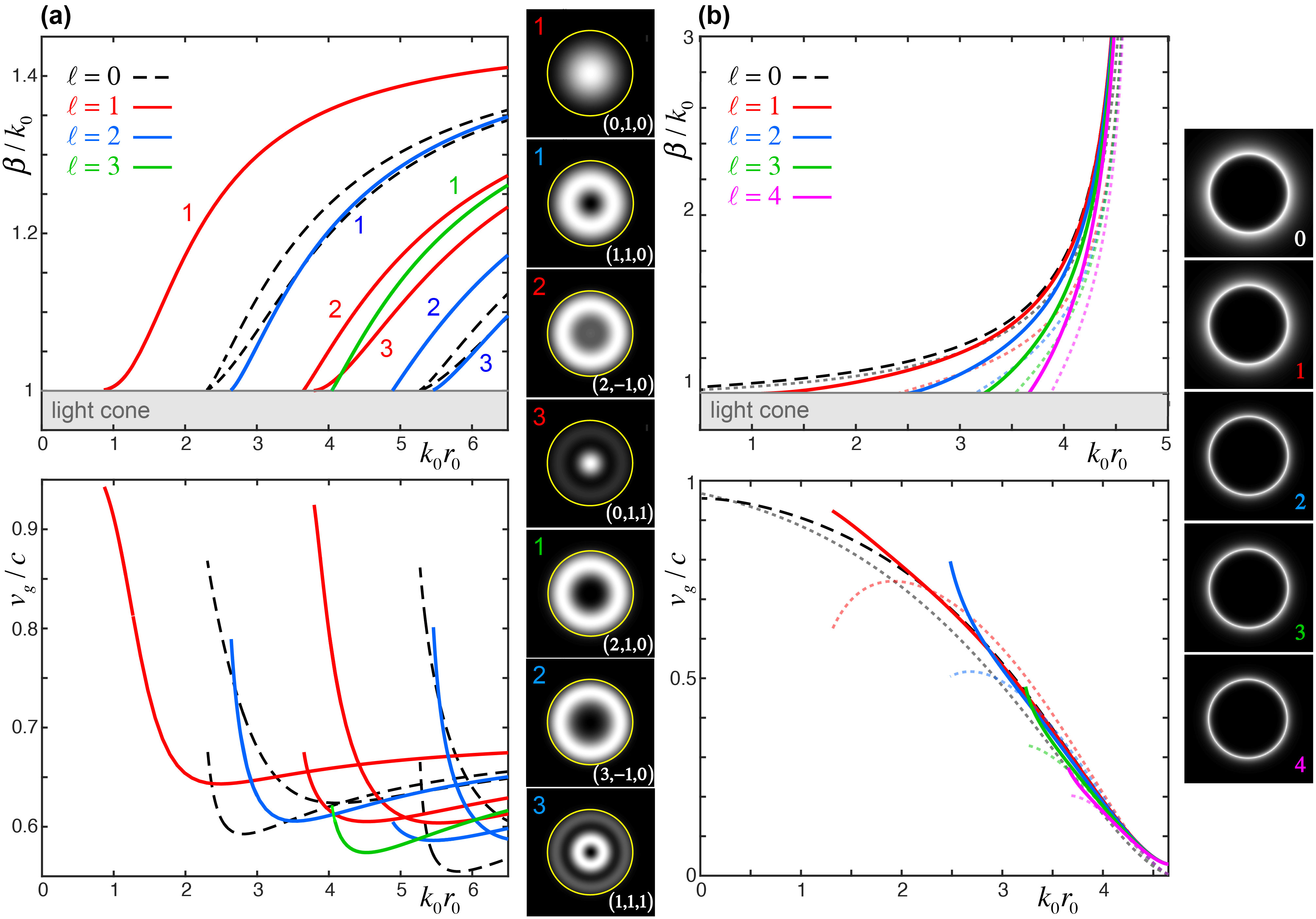}}
\caption{Numerically calculated eigenmodes of a multimode dielectric fiber with parameters $r_0=200$ nm, $\varepsilon_1 = 2.1$, $\varepsilon_2 = 1$ (a) and of a metallic wire with parameters $r_0 = 150$ nm, $\varepsilon_1 = 1-\omega^2_p /\omega^2$, $\omega_p=1.3262\times 10^{16}\,{\rm s}^{-1}\simeq 6.63~c/r_0$, $\varepsilon_2 = 1$ (b). The frequency $\omega$ was varied in these calculations. The upper panels depict the normalized propagation constants $\beta$ , which characterize the canonical momentum (\ref{eq:group_vel}) of the modes (exceeding $\hbar k_0$ per photon). The lower panels show the subluminal group velocities (\ref{eq:group_vel}) of the modes. The small greyscale panels show typical transverse energy distributions $W(x,y)$ in different modes. The dielectric fiber modes are marked by the total-AM quantum number $\ell = m +\sigma$, as well as by the three (orbital, spin, and radial) quantum numbers $(m,\sigma,n)$ Eq.~(\ref{eq:quantum_numbers}). The metallic-wire modes are marked by the single total-AM quantum number $\ell$. The dotted curves in (b) correspond to the surface-plasmon geometrical-optics model, Eqs.~(\ref{eq:an_metal_modes}) and (\ref{eq:metal_gv}).}
\label{fig:fig1}
\end{figure*}

%%%%%%%%%%%%%%%%%%%%%%%%%%%%%%%%%%%%%%%%%%
%
The transcendental characteristic equation for $\beta$ and cumbersome relations for the constants $(A,B,C,D)$ require numerical calculations. Figure~\ref{fig:fig1} shows examples of the numerically-calculated dispersions $\beta(\omega)$ and energy distributions $W(x,y)$ for the eigenmodes of multimode dielectric fibers and metallic wires. These modes can be classified via their \emph{quantum numbers}. As we show below, the main azimuthal quantum number $\ell$ characterizes the \emph{total AM}. The $\ell = 0$ modes are pure TE (with $A=C=0$) and TM waves (with $B =D =0$) (see Appendix~A) \cite{marcuse1972light}, for which the AM and helicity vanish identically:
\begin{equation}
\label{eq:ell0}
L_z=S_z = \mathfrak{S}=J_z=\mathcal{J}_z=0 \quad \text{for}\quad \ell = 0.
\end{equation}
Therefore, in what follows, we are interested only in the $\ell \neq 0$ modes, which are \emph{mixed} (i.e., neither TE, nor TM). 
	
%%%%%%%%%%%%%%%%%%%%%%%%%%%%%%%%%%%%%%%%%%
%%%%%%%%%%%%%%%%%%%%%%%%%%%%%%%%%%%%%%%%%%

Importantly, in dielectric fibers, these modes (including the fundamental mode with $\ell =1$) have circular polarizations in the paraxial limit \cite{snyder2012optical,dooghin1992optical}. This corresponds to geometrical-optics rays propagating inside the dielectric due to the total internal reflection and having circular polarizations, as shown in Fig~\ref{fig:fig0}(a) \cite{snyder2012optical}. Therefore, one can introduce the \emph{spin quantum number} $\sigma = \pm 1$, characterizing the sign of this polarization, spin AM, and helicity of the mode in the paraxial limit (where the geometrical-optics rays are practically aligned with the $z$-axis). Accordingly, the orbital AM of the mode in the paraxial approximation is described by \emph{the orbital AM quantum number} $m=\ell - \sigma$, which corresponds to the orbital AM carried by helical geometrical-optics rays, Fig.~\ref{fig:fig0}(a) \cite{snyder2012optical}. Finally, the dielectric-fiber modes with the same AM numbers can have different radial profiles, which are characterized by the \emph{radial quantum number} $n = 0,1,2,...$, counting the number of additional maxima of $W(r)$ and corresponding to the fact that the geometrical-optics rays can propagate at different angles with respect to the dielectric interface \cite{snyder2012optical}. Thus, a set of three quantum numbers,
\begin{equation}
\label{eq:quantum_numbers}
(m,\sigma,n) = (\text{orbital, spin, radial}),
\end{equation}
labels the $\ell \neq 0$ modes of a dielectric fiber, as shown in Fig~\ref{fig:fig1}(a) \cite{snyder2012optical,dooghin1992optical}. Due to the mirror symmetry of the waveguide, the modes with opposite total AM $\ell = \pm 1, \pm 2, ...$ are double-degenerate, and we restrict our analysis to the $\ell > 0$ case. At the same time, the modes with opposite spin quantum numbers $\sigma = \pm 1$ (and the same orbital and radial quantum numbers) are not degenerate, which indicates the \emph{spin-orbit interaction} in optical fibers \cite{dooghin1992optical,liberman1992spin,bliokh2004modified,alexeyev2007fiber, golowich2014asymptotic,bliokh2015spin,bliokh2008geometrodynamics}.

The situation is much simpler in the case of metallic wires. There, the eigenmodes have a \emph{surface plasmon-polariton} origin \cite{pfeiffer1974surface,novotny1994light,maier2007plasmonics,novotny2012principles}. Therefore, the mode is localized near the metal-dielectric interface [all geometrical-optics rays lie on the cylindrical surface, Fig.~\ref{fig:fig0}(b)], and its radial profile is fixed for each $\ell$, i.e., effectively $n\equiv 0$. Furthermore, the polarization is also fixed, locally tending to the TM surface-plasmon mode in the large-radius limit $k_0r_0 \gg 1$; i.e., there are no circularly-polarized modes and effectively $\sigma\equiv 0$. Thus, the metallic-wire modes are labelled by a single \emph{total AM quantum number} $\ell$, as shown in Fig.~\ref{fig:fig1}(b). 

Nonetheless, the AM and helicity properties of the metallic-wire modes are generally nontrivial. Akin to the dielectric-fiber case, the fundamental $\ell=0$ mode has pure TM polarization with $B=D=0$ and vanishing AM and helicity, Eq.~(\ref{eq:ell0}). However, the higher-order modes are mixed, and, as we show below, their spin and orbital AM, as well as helicity, are nonzero. Notably, the nonzero spin AM of the metallic-wire modes can be explained by the fact that even locally-TM-polarized surface-plasmon waves possess an elliptical polarization in the propagation plane [see Fig.~\ref{fig:fig0}(b)] and therefore carries the {\it transverse spin} \cite{bliokh2017optical,bliokh2017optical2,bliokh2012transverse}, a phenomenon which is currently attracting considerable attention
\cite{bliokh2015transverse,bliokh2015spin,aiello2015transverse}. 
For the modes with $\ell > 0$, the geometrical-optics surface-plasmon rays are \emph{helical} \cite{catrysse2009understanding}, Fig.~\ref{fig:fig0}(b), and the locally-transverse spin acquires a nonzero $z$ component. In Section~\ref{sec:explicit}\ref{subsec:metal_wires}, we will show that this geometrical-optics ray picture, supplied by the known surface-plasmon-polariton properties, describes properties of higher-order metallic-wire modes and enables one to derive approximate analytical expressions for the dispersion and AM quantities. To the best of our knowledge, the nonzero spin AM of higher-order metallic-wire modes and its relation to the transverse spin of surface plasmon-polaritons has never been described before.

%%%%%%%%%%%%%%%%%%%%%%%%%%%%%%%%%%%%%%%%%%
\section{Angular momenta and momenta of guided modes}
\label{sec:angular_momenta}
%%%%%%%%%%%%%%%%%%%%%%%%%%%%%%%%%%%%%%%%%%
\subsection{Momentum, spin, orbital, and total angular momenta}
\label{subsec:momenta}
%%%%%%%%%%%%%%%%%%%%%%%%%%%%%%%%%%%%%%%%%%
%
Some important momentum and AM properties of the cylindrical modes can be found analytically from Eqs.~(\ref{eq:eigenmode}) and (\ref{eq:eigenmode_out}), without numerically calculating their parameters. In this section, we describe these universal momentum and AM features, independent of the dielectric or metallic waveguide properties. We first note that all field components (\ref{eq:eigenmode}) share the same $z$-dependent factor exp($i\beta z$). From here, it is easy to see that the $z$-component of the canonical momentum (\ref{eq:definitions}) is naturally associated with the propagation constant of the mode, $\beta$. At the same time, the integral Poynting vector (\ref{eq:poynting}) provides the group velocity of the modes \cite{snyder2012optical,marcuse1972light,jackson1999classical}.
These momentum and velocity properties read:
\begin{equation}\label{eq:group_vel}
\frac{\left\langle P_z \right\rangle}{\left\langle W \right\rangle}=\frac{P_z}{W}=\frac{\beta}{\omega}, \quad v_g = \frac{c^2\left\langle \mathcal{P}_z \right\rangle}{\left\langle W \right\rangle}=\frac{\partial \omega}{\partial \beta},
\end{equation}
where $\left\langle...\right\rangle$ denotes the integration over the transverse ($x$,$y$)-plane. Note that since $\beta >k_0$, the canonical momentum per photon always exceeds the photon momentum in vacuum. In other words, the guided modes carry “\emph{supermomentum}” larger than $\hbar k_0$ per photon \cite{bliokh2017optical,bliokh2017optical2,berry2009optical,bliokh2013photon, barnett2013superweak}. 
At the same time, the group velocity is always subluminal: $v_g<c$. This imposes the following inequality on the Poynting and canonical momenta: $c \left\langle P_z \right\rangle/\left\langle W \right\rangle > 1 > c \left\langle \mathcal{P}_z\right\rangle/\left\langle W \right\rangle$,  which seem to
be universal for any guided modes \cite{bliokh2017optical,bliokh2017optical2}, while for free-space localized solutions $c \left\langle P_z \right\rangle/\left\langle W \right\rangle=c\left\langle\mathcal{P}_z\right\rangle/\left\langle W \right\rangle<1$ \cite{bliokh2013dual,bliokh2013photon}. Figure~\ref{fig:fig1} shows these dimensionless canonical-momentum and group-velocity characteristics for the numerically-calculated modes of dielectric fibers and metallic wires, confirming that these are restricted by 1 from below and above, respectively.

The eigenmodes fields (\ref{eq:eigenmode}) and (\ref{eq:eigenmode_out}) are written in a form convenient for the AM analysis. Indeed, each field component has a well-defined vortex phase factor exp($i\alpha\phi$). In turn, the $z$-component of the orbital AM (\ref{eq:definitions}) is determined by the operator $\hat{L}_z=-i(\mathbf{r}\times \nabla)_z=-i\,\partial /\partial\phi$. However, the whole field (\ref{eq:eigenmode}) is not an orbital AM eigenmode, because different components
have different azimuthal numbers $\alpha$. This is typical for nonparaxial vortex fields with intrinsic spin-orbit coupling \cite{van1994commutation,bliokh2010angular,bliokh2015spin}.

For the analysis of the AM properties of the modes, it is instructive to write the energy density (\ref{eq:definitions}) as a sum of the energies of the right-hand circular ($+$), left hand circular ($-$), and longitudinal ($z$) field components: $W=W^+ + W^- + W_z$, where $W^\pm = \left(\tilde{\varepsilon}\left|E^\pm\right|^2+\tilde{\mu}\left|H^\pm\right|^2\right)\!/4$ and $W_z = \left(\tilde{\varepsilon}\left|E_z\right|^2+\tilde{\mu}\left|H_z\right|^2\right)\!/4$. Substituting now the fields (\ref{eq:eigenmode}) and 
(\ref{eq:eigenmode_out}) into Eqs.~(\ref{eq:definitions}), we find that
the $z$-components of the spin and orbital AM can be written as:
\begin{eqnarray}
\label{eq:z_comp}
&&\frac{\omega L_z}{W}=\frac{(\ell -1)W^+\! + (\ell +1)W^-\! + \ell\, W_z}{W},
\nonumber\\
&&\frac{w S_z}{W} = \frac{W^+\! - W^-}{W}.
\end{eqnarray}
Most importantly, it follows from these relations that the total AM of the eigenmodes is always an \emph{integer}:
\begin{equation}\label{eq:integer}
\frac{\omega\left\langle J_z \right\rangle}{\left\langle W \right\rangle} = \frac{\omega J_z}{W} = \frac{\omega L_z}{W}+\frac{\omega S_z}{W}=\ell.
\end{equation}

To the best of our knowledge, this remarkably simple result has not been derived before.
Moreover, it is by no means trivial. On the one hand, a cylindrically-symmetric stationary system must possess eigenmodes, simultaneously, of the energy ($i\partial/\partial t$) and total AM ($\hat{J}_z$) operators, with the corresponding eigenvalues $\omega$ and $\ell$. On the other hand, until recently, we have not had expressions for the total AM of light in a medium, which would yield the integer value (\ref{eq:integer}). In particular, the often-used Poynting-Abraham total AM (\ref{eq:kinetic_am_desity}) is not an integer for cylindrical guided waves (see \cite{le2006angular} and Figs.~\ref{fig:fig2} and \ref{fig:fig3} below):
\begin{equation}\label{eq:not_integer}
\frac{\omega \left\langle\mathcal{J}_z\right\rangle}{\left\langle W \right\rangle}\neq \ell.
\end{equation}
It is only the recently-derived canonical Minkowski-type AM \cite{bliokh2017optical2,bliokh2017optical} that yields the proper integer value (\ref{eq:integer}). We also emphasize the importance of the \emph{dual-symmetric} form of the canonical energy, momentum, and AM expressions (\ref{eq:definitions}), which can be written as a sum of the electric and magnetic contributions: $\mathbf{P}=\mathbf{P}^e + \mathbf{P}^m$, $\mathbf{L}=\mathbf{L}^e + \mathbf{L}^m$, $\mathbf{S}=\mathbf{S}^e + \mathbf{S}^m$. The simple results (\ref{eq:group_vel})–(\ref{eq:integer}) would not be obtained for the {\it pure-electric} definitions $\mathbf{P}'=2\mathbf{P}^e$, $\mathbf{L}'=2\mathbf{L}^e$, $\mathbf{S}'=2\mathbf{S}^e$. Obtaining the values (\ref{eq:group_vel})–(\ref{eq:integer}) for the electric-biased definitions would require to also use the pure-electric energy $W'=2W^e$, as was done in~\cite{le2006angular}. However, such definition is physically inconsistent because the pure-electric energy is not a conserved quantity, even in free space. The fundamental importance and consistency of the canonical Minkowski-type dual-symmetric definitions (\ref{eq:definitions}) is discussed in detail in \cite{bliokh2017optical2,bliokh2017optical}. The natural and laconic form of Eqs.~(\ref{eq:group_vel})–(\ref{eq:integer}) fairly supports this approach.

As we will see in Section~\ref{sec:explicit}, the dielectric-fiber modes become paraxial and circularly-polarized, with $\omega\left\langle S_z \right\rangle/\left\langle W \right\rangle \simeq\omega \left\langle\mathfrak{S}\right\rangle/\left\langle W \right\rangle \simeq \sigma = \pm 1$ and $\omega\left\langle L_z \right\rangle/\left\langle W \right\rangle \simeq m= \ell - \sigma$ in the $k_0r_0\gg 1$ limit. This determines the spin and orbital quantum numbers (\ref{eq:quantum_numbers}). In the nonparaxial regime, these values are not integer but the sign of the spin AM and helicity still determines the quantum number $\sigma$. For the metallic-wire modes, $\omega\left\langle S_z \right\rangle/\left\langle W \right\rangle \simeq \omega\left\langle\mathcal{S}\right\rangle/\left\langle W \right\rangle \simeq 0$ in the $k_0r_0\gg 1$ limit. Note also that the vanishing spin and orbital AM of pure TE and TM modes with $\ell=0$, Eq.~(\ref{eq:ell0}), follows from Eqs.~(\ref{eq:z_comp}), (\ref{eq:integer}) and Eqs.~(\ref{eq:eigenmode}), (\ref{eq:eigenmode_out}) with $A = C = 0$ or $B = D = 0$, when we notice that $\left|E^+\right|^2=\left|E^-\right|^2$, $\left|H^+\right|^2=\left|H^-\right|^2$, and hence $W^+ = W^-$.

%%%%%%%%%%%%%%%%%%%%%%%%%%%%%%%%%%%%%%%%%%
\subsection{Relation to the dynamical and geometric phases}
%%%%%%%%%%%%%%%%%%%%%%%%%%%%%%%%%%%%%%%%%%
%
Remarkably, the values of the angular momenta (\ref{eq:z_comp}) and (\ref{eq:integer}), as well as the quantization of the total AM, are closely related to the \emph{dynamical and geometric phases} in inhomogeneous polarized fields. 

To start with, we would like to characterize the phase difference in a complex vector field $\boldsymbol{\psi}(\mathbf{r})$ between two $\mathbf{r}$-points connected by a contour $\mathcal{C}$. For a scalar field $\psi(\mathbf{r})$, the only natural definition of the phase is $\Phi = \int_\mathcal{C} {{\nabla}\text{Arg}(\psi)\cdot d\mathbf{r}}=\mathrm{Im}\int_\mathcal{C} {\frac{\psi^{*}(\nabla)\psi}{\left|\psi\right|^2}\cdot d\mathbf{r}}$. However, the vector field $\boldsymbol{\psi}(\mathbf{r})$ has more degrees of freedom: for example, it can be factorized into a complex scalar amplitude and a unit direction (polarization) vector. One way to introduce the phase is to use the scalar complex field $\Psi = \boldsymbol{\psi}\cdot \boldsymbol{\psi}$ \cite{nye1991phase,berry2001polarization}:
\begin{equation}
\label{eq:dynamical_phase}
\Phi_D=\frac{1}{2}\int_\mathcal{C} {\nabla\text{Arg}(\Psi)\cdot d\mathbf{r}}.
\end{equation}
This phase can be associated with the {\it dynamical} phase in the field, because it is independent of the direction of the field polarization. Alternatively, one can calculate the phase using the local wavevector of the field, determined by the expectation value of the $-i\nabla$ (canonical-momentum) operator \cite{berry2009optical,bliokh2013photon,barnett2013superweak}:
\begin{equation}\label{eq:tot_phase}
\Phi = \int_\mathcal{C} {\mathbf{k}_{\rm loc}\cdot d\mathbf{r}}\equiv\mathrm{Im}\int_\mathcal{C} {\frac{\boldsymbol{\psi}^*\cdot (\nabla)\boldsymbol{\psi}}{\boldsymbol{\psi}^*\cdot \boldsymbol{\psi}}\cdot d\mathbf{r}}.
\end{equation}
This phase can be called the {\it total} phase of the field, because the operator $-i\nabla$ acts on both the scalar and polarization parts of the vector field. Accordingly, the difference between the phases (\ref{eq:tot_phase}) and (\ref{eq:dynamical_phase}) is the {\it geometric phase} caused by the inhomogeneous polarization along the contour $\mathcal{C}$:
\begin{equation}
\label{eq:geom_phase}
\Phi_G = \Phi - \Phi_D.
\end{equation}
We analyze this phase in detail elsewhere \cite{bliokh2018inpreparation}; in particular, we show that it coincides with the well-known Pancharatnam-Berry phase on the Poincaré sphere \cite{bhandari1997polarization} in the case of paraxial fields.

To apply this formalism to the electromagnetic field in optical media, we introduce the 6-component electromagnetic “wavefunction” $\boldsymbol{\psi}=\omega^{-1/2}(\mathbf{E},\mathbf{H})$. Importantly, the scalar product for this Maxwell field in a dispersive inhomogeneous medium should be modified, because the macroscopic Maxwell equations are effectively non-Hermitian. As it was shown recently \cite{alpeggiani2018electromagnetic} (see also \cite{gangaraj2017berry,de2017schr}), the modified inner product in a medium involves the “left vector” $\tilde{\boldsymbol{\psi}}=\omega^{-1/2}(\tilde{\varepsilon}\,\mathbf{E},\tilde{\mu}\,\mathbf{H})$, i.e., 
$\boldsymbol{\psi}^*\!\cdot(...)\boldsymbol{\psi}\rightarrow\tilde{\boldsymbol{\psi}}^*\!\cdot(...)\boldsymbol{\psi}$. 
With this modified scalar product, the canonical momentum, spin, and orbital AM (\ref{eq:definitions}), as well as helicity (\ref{eq:helicity}) represent the local
expectation values of the corresponding quantum operators \cite{bliokh2017optical2,bliokh2017optical,alpeggiani2018electromagnetic}, while the Brillouin energy density is determined by the wavefunction norm: $W = \omega\, \tilde{\boldsymbol{\psi}}^*\!\cdot \boldsymbol{\psi}$. 
Furthermore, substituting the “right” and “left” electromagnetic wavefunctions into Eqs.~(\ref{eq:dynamical_phase})–(\ref{eq:geom_phase}), we can now calculate the increments of the phases (\ref{eq:dynamical_phase})–(\ref{eq:geom_phase}) for the waveguide modes (\ref{eq:eigenmode}) and (\ref{eq:eigenmode_out}) along a closed circuit $\mathcal{C} = \{r=\text{const},\phi\in (0,2\pi)\}$. In doing so, $\nabla \cdot d\mathbf{r} = (\partial/\partial \phi)\,d \phi = i\, \hat{L}_z\, d\phi$, and the total phase increment (\ref{eq:tot_phase}) becomes naturally proportional to the canonical orbital AM (\ref{eq:z_comp}). Moreover, the dynamical phase (\ref{eq:dynamical_phase}) becomes proportional to the total AM (\ref{eq:integer}), while the geometric phase (\ref{eq:geom_phase}) becomes proportional to the minus spin AM:
\begin{equation}\label{eq:phase_bar}
\frac{\omega L_z}{W}=\stkout{\Phi}, \quad \frac{\omega S_z}{W} = - \stkout{\Phi}_G, \quad \frac{\omega J_z}{W}=\stkout{\Phi}_D=\ell,
\end{equation}
where $\stkout{\Phi}=\Phi/2\pi$. The last equality in Eq.~(\ref{eq:phase_bar}) readily follows from the definition (\ref{eq:dynamical_phase}) and fields (\ref{eq:eigenmode}) if we notice that for the circular-polarized components $\boldsymbol{\psi}\cdot\boldsymbol{\psi}=2\psi^+ \psi^- + \psi_z^2 \propto \text{exp}(2i\ell\phi)$. Thus, the quantization of the total AM is explained by the quantization of the dynamical phase along the circuit $\mathcal{C}$ (this characterizes the topological vortex number of the scalar field $\Psi = \boldsymbol{\psi}\cdot \boldsymbol{\psi}$). The proportionality between the spin AM and geometric phase is also easy to explain. Moving along the contour $\mathcal{C}$, we are attached to the cylindrical coordinates ($r$,$\phi$) which experience a $2\pi$ rotation with respect to the Cartesian axes $(x,y)$. Therefore, the right-hand ($+$) and left-hand ($-$) circular field components acquire the opposite geometric phases $\mp 2\pi$ \cite{bliokh2015spin,bliokh2008coriolis}, which are averaged in the second Eq.~(\ref{eq:z_comp}) with the weights $W^+$ and $W^-$.

These results resemble previous calculations of the spin and orbital AM in nonparaxial
Bessel beams in free space \cite{bliokh2010angular,bliokh2014conservation}. However, there are two differences. First, most importantly, the free-space consideration \cite{bliokh2010angular} is based on the \emph{Fourier plane-wave decomposition} of the field and the spin-redirection geometric phase in $\mathbf{k}$-\textit{space}. In the present problem, this approach is inapplicable because plane waves are \emph{not} eigenmodes of an inhomogeneous cylindrical medium. Therefore, our treatment is based on another type of geometric phase (similar to the Pancharatnam-Berry one) in $\mathbf{r}$-\textit{space} \cite{bliokh2018inpreparation}. Second, one can notice the difference between Eqs.~(\ref{eq:phase_bar}) and analogous equations in Ref.~\cite{bliokh2010angular}. This is because the free-space Bessel beams in \cite{bliokh2010angular} are defined such that $\ell$ is the \textit{orbital-AM} number (corresponding to $m$ in this work), and the $\ell = 0$ beam tends to a uniform circularly polarized plane wave in the paraxial limit. In contrast, the cylindrical-waveguide modes (\ref{eq:eigenmode}) and (\ref{eq:eigenmode_out}) are defined with respect to the polar coordinates, so that $\ell$ is the \textit{total-AM} quantum number, and the $\ell = 0$ modes are singular on axis ($r = 0$). The two approaches are connected by the substitution $\ell = m +\sigma$, where $\sigma = \pm 1$ is the spin/helicity quantum number. Making this substitution in Eqs.~(\ref{eq:z_comp}), (\ref{eq:integer}), and (\ref{eq:phase_bar}), we find that the spin, orbital, and total AM could be written as $\omega L_z /W = m + \stkout{\Phi}'_G$, $\omega S_z /W = \sigma -\stkout{\Phi}'_G$, $\omega J_z/W = m +\sigma$, where the modified geometric phase (now defined with respect to the Cartesian rather than polar axes) is $\Phi'_G=\Phi_G +2\pi\sigma$. These relations have exactly the same form as the ones derived for the free-space Bessel beams \cite{bliokh2010angular}.

%%%%%%%%%%%%%%%%%%%%%%%%%%%%%%%%%%%%%%%%%%
\section{Explicit calculations}
\label{sec:explicit}
%%%%%%%%%%%%%%%%%%%%%%%%%%%%%%%%%%%%%%%%%%
\subsection{Dielectric fibers}
%%%%%%%%%%%%%%%%%%%%%%%%%%%%%%%%%%%%%%%%%%
%
We are now in a position to show explicit results for the AM and helicity values for the cylindrical guided modes. We first consider dielectric fibers, which are assumed to be made of nondispersive materials: $\tilde{\varepsilon}=\varepsilon$ and $\tilde{\mu}=\mu$.

%%%%%%%%%%%%%%%%%%%%%%%%%%%%%%%%%%%%%%%%%%
\begin{figure*}[t]
\centering
\fbox{\includegraphics[width=0.99\linewidth]{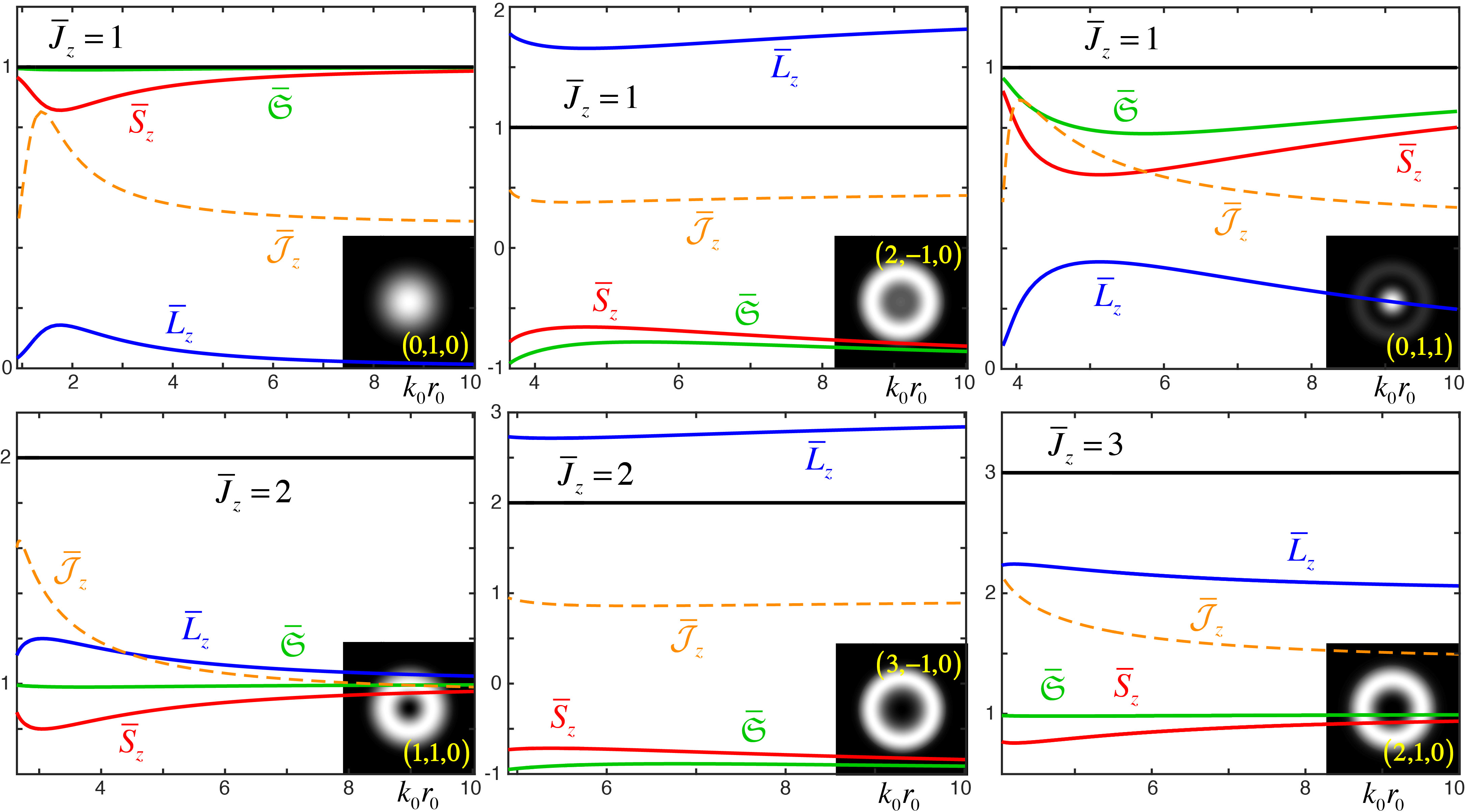}}
\caption{Numerically calculated canonical spin, orbital and total AM [Eqs.~(\ref{eq:definitions}), (\ref{eq:z_comp}), (\ref{eq:integer}), and (\ref{eq:dielectric_simp})] as well as the helicity [Eqs.~(\ref{eq:helicity}) and (\ref{eq:dielectric_simp})] and the Abraham-Poynting total AM [Eq.~(\ref{eq:kinetic_am_desity})] of the modes of a dielectric fiber shown in Fig.~\ref{fig:fig1}(a).Here, plotted are the normalized integral values (in units of $\hbar$ per photon), defined as $\bar{...}=\omega \left\langle ... \right\rangle/\left\langle W \right\rangle$. One can see the quantization of the canonical total AM $\overline{J}_z = \overline{L}_z+\overline{S}_z = \ell$, the non-integer Poynting-Abraham AM $\overline{\mathcal{J}}_z \neq \ell$, and the differing spin AM and helicity $\overline{S}_z\neq\overline{\mathfrak{S}}$. In the large-radius (paraxial) limit $k_0r_0\gg 1$, the canonical spin and orbital AM tend to the quantized values $\overline{L}_z\simeq m$ and $\overline{S}_z\simeq \overline{\mathfrak{S}} \simeq \sigma$.}
\label{fig:fig2}
\end{figure*}
%%%%%%%%%%%%%%%%%%%%%%%%%%%%%%%%%%%%%%%%%%

Apart from the general result for the canonical total AM (\ref{eq:integer}), the fields, dispersion, and dynamical properties of the modes require numerical calculations. These can be performed
directly using the equations of Section~\ref{sec:basic} and Appendix A. However, we found that a considerable analytical simplification can be executed. Namely, substituting Eq.~(\ref{eq:eigenmode}) into Eqs.~(\ref{eq:definitions}), (\ref{eq:poynting}), and (\ref{eq:helicity}), we derive the following expressions for the energy, spin, helicity, and Poynting momentum densities inside the fiber $(r<r_0)$:
\begin{eqnarray}
\label{eq:dielectric_simp}
W\!=\!\frac{1}{4}[b\xi^- G + (a \xi^+ + \zeta)F ], ~~ 
\mathcal{P}_z\!=\!\frac{1}{4\sqrt{\varepsilon\mu}c}[b \xi^+ F + a\xi^- G],\nonumber\\
S_z\!=\!\frac{1}{4\omega}[a\xi^- F + b \xi^+ G], ~~ 
\mathfrak{S}\!=\!\frac{1}{4\omega}[b\xi^- F + (a\xi^+ + \zeta)G].
\end{eqnarray}
Here, we introduced the following parameters:
\begin{eqnarray}\label{eq:dielectric_param}
\xi^\pm(\rho)=\left|J_{\ell-1}(\rho)\right|^2\pm \left|J_{\ell+1}(\rho)\right|^2, \quad \zeta(\rho) = 2\left|J_{\ell}(\rho)\right|^2,\nonumber\\
a = \frac{k^2+\beta^2}{|\kappa|^2}, ~~ b=\frac{2 k\beta}{|\kappa|^2}, ~~
 F = |A|^2+|B|^2, ~~ G=2\,\mathrm{Im}(AB^*).
\end{eqnarray}
Outside of the fiber $(r>r_0)$, the energy, helicity, and spin densities are given by Eqs.~(\ref{eq:dielectric_simp}) and (\ref{eq:dielectric_param}) with the substitution (\ref{eq:eigenmode_out}). Note that the canonical momentum and the orbital or total AM do not require additional calculations, because, according to Eqs.~(\ref{eq:group_vel}) and (\ref{eq:integer}), they are determined by the energy and spin densities: $P_z = \beta W/\omega$, $J_z=\ell W/\omega$, and $L_z = J_z-S_z$.

Equations (\ref{eq:dielectric_simp}) and (\ref{eq:dielectric_param}) illuminate some properties of the spin and helicity in the waveguide modes, clearly showing that these are different quantities, which characterize the intrinsic AM \cite{allen2003optical,torres2011twisted,andrews2012angular,allen1999iv,yao2011orbital,bliokh2015transverse,berestetskii1982quantum,soper2008classical,leader2016photon,van1994commutation,bliokh2010angular,barnett2010rotation,bialynicki2011canonical,bliokh2014conservation} and chirality of the field \cite{afanasiev1996helicity,trueba1996electromagnetic,cameron2012optical,bliokh2013dual,fernandez2013electromagnetic,van2016conditions,alpeggiani2018electromagnetic}, respectively. First, the helicity coincides with the $z$-component of the spin AM only in the paraxial limit. Indeed, the paraxial limit $\kappa \ll k$ corresponds to $b\simeq a\gg 1$, and $\mathfrak{S}\simeq S_z$. Second, it is easy to see that $F\geq |G|$, and the helicity magnitude is restricted by the fundamental limit of 1 (in $\hbar$ units per photon): $\omega|\mathfrak{S}|/W\leq 1$. Third, the helicity eigenstates with $\omega|\mathfrak{S}|/W=1$ correspond to $\mathfrak{S}=\pm W$, $F=\pm G$, which yields $A=\pm i B$ or $C= \pm i D$. This condition means that the fields (\ref{eq:eigenmode}) and (\ref{eq:eigenmode_out}) satisfy $\mathbf{E}=\pm i \sqrt{\frac{\mu}{\varepsilon}}\mathbf{H}$, which are exactly the eigenstates of the helicity operator in a medium: $\hat{\mathfrak{S}}=\left(\begin{matrix}
0 && i\sqrt{\mu/\varepsilon}\\
-i\sqrt{\varepsilon/\mu} && 0
\end{matrix}\right)$ acting on the vector 
$\boldsymbol{\Psi}\propto \left(\begin{matrix}
\mathbf{E}\\
\mathbf{H}
\end{matrix}\right)$ 
\cite{fernandez2013electromagnetic,van2016conditions,alpeggiani2018electromagnetic}. Finally, the helicity and longitudinal spin of the fields (\ref{eq:eigenmode}) and (\ref{eq:eigenmode_out}) are nonzero in the general case, because these are mixed (i.e., neither TE nor TM) modes. The only exception is the $\ell=0$ case, where, for the TE ($A=C=0$) and TM ($B=D=0$) modes, we have $\xi^-(\rho)=G=0$, and all helicity and AM properties vanish in agreement with Eq.~(\ref{eq:ell0}).\\

Figure~\ref{fig:fig2} shows the results of numerical calculations of the integral values of the spin/orbital/total AM and helicity, $\left\langle S_z\right\rangle$, $\left\langle L_z\right\rangle$, $\left\langle \mathcal{J}_z\right\rangle$, and $\left\langle \mathfrak{S}\right\rangle$ for several dielectric-fiber modes shown in Fig.~\ref{fig:fig1}(a). One can clearly see the quantization of the canonical total AM, noninteger character of the Poynting-Abraham total AM, and helicity different from the spin. While $\ell$ is the total AM quantum number, these calculations allow one to identify the spin and orbital quantum numbers, $\sigma = \text{sgn}\left\langle S_z\right\rangle$ and $m = \ell-\sigma$, discussed in Section \ref{sec:basic}\ref{subsec:labelling}. One can also see that the normalized spin/helicity and orbital AM values (but not the Poynting-Abraham AM) tend to:
\begin{equation}
\label{eq:parax_dielec}
\frac{\omega\left\langle S_z \right\rangle}{\left\langle W \right\rangle}\simeq \frac{\omega\left\langle\mathfrak{S}\right\rangle}{\left\langle W \right\rangle}\simeq\sigma, \quad \frac{\omega\left\langle L_z\right\rangle}{\left\langle W \right\rangle}\simeq m, \quad \text{for}\quad k_0r_0\ll 1.
\end{equation}
The non-integer character of these quantities in the general nonparaxial case signals the spin-orbit interaction of light in the fiber \cite{bliokh2010angular,bliokh2014conservation,dooghin1992optical,liberman1992spin, bliokh2004modified,alexeyev2007fiber,golowich2014asymptotic,bliokh2015spin}.

%%%%%%%%%%%%%%%%%%%%%%%%%%%%%%%%%%%%%%%%%%
\begin{figure*}[t]
\centering
\fbox{\includegraphics[width=0.99\linewidth]{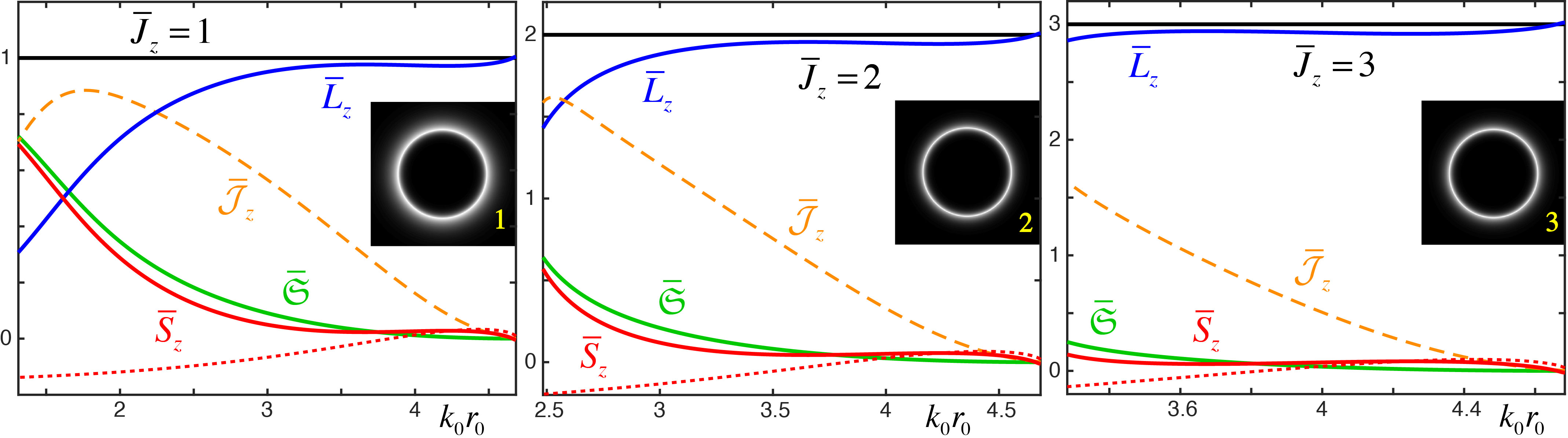}}
\caption{Same as in Fig.~\ref{fig:fig2} but for the metallic-wire modes shown in Fig.~\ref{fig:fig1}(b). The main difference in the behavior of the depicted quantities is that in the large-radius (paraxial) limit $k_0r_0\gg 1$, the canonical spin and orbital AM tend to the values $\overline{L}_z\simeq\ell$ and $\overline{S}_z\simeq\overline{\mathfrak{S}}\simeq 0$, whereas, surprisingly, the Poynting-Abraham total AM also vanishes: $\overline{\mathcal{J}}_z \simeq 0$. The red dotted curves correspond to the geometrical-optics model for the spin AM, Eq.~(\ref{eq:long_spin}), based on the transverse spin of surface plasmon-polaritons.}
\label{fig:fig3}
\end{figure*}
%%%%%%%%%%%%%%%%%%%%%%%%%%%%%%%%%%%%%%%%%%

%%%%%%%%%%%%%%%%%%%%%%%%%%%%%%%%%%%%%%%%%%
\subsection{Metallic wires}\label{subsec:metal_wires}
%%%%%%%%%%%%%%%%%%%%%%%%%%%%%%%%%%%%%%%%%%
%
We now consider cylindrical metallic wires characterized by the dispersive permittivity $\varepsilon_1(\omega)<0$ and the corresponding $\tilde{\varepsilon}_1 >0$. Figure~\ref{fig:fig3} shows calculations analogous to Fig.~\ref{fig:fig2}, using the general equations of Section~\ref{sec:basic}, but now for the eigenmodes of a metallic wire, Fig.~\ref{fig:fig1}(b). There is one important difference in the behavior of the spin and helicity in Figs.~\ref{fig:fig2} and \ref{fig:fig3}. Namely, in the paraxial (large-radius) limit, the metallic-wire modes tend to the TM surface plasmon-polariton waves (the wire surface can be locally approximated by a planar interface)
with vanishing longitudinal spin and helicity. Moreover, surprisingly, the Poynting-Abraham total AM also vanishes in this limit:
\begin{equation}
\label{eq:parax_metal}
\frac{\omega\left\langle S_z \right\rangle}{\left\langle W \right\rangle}\simeq \frac{\omega\left\langle\mathfrak{S}\right\rangle}{\left\langle W \right\rangle}\simeq\frac{\omega\left\langle \mathcal{J}_z \right\rangle}{\left\langle W \right\rangle}\simeq 0,~~ 
\frac{\omega\left\langle L_z\right\rangle}{\left\langle W \right\rangle}\simeq \ell, ~~ \text{for}~~ k_0r_0\ll 1.
\end{equation}
The vanishing Poynting-Abraham AM exhibits a dramatic difference with the quantized canonical AM. This behavior can be qualitatively explained as follows. In the large-radius limit, the mode is locally described by the near-planar surface plasmon-polariton wave propagating at an angle with respect to the $z$-axis, Fig.~\ref{fig:fig0}(b). The Poynting vector of this surface plasmon-polariton has a nonzero azimuthal component $\mathcal{P}_\phi$, which determines the $z$-component of the Poynting-Abraham total AM: $\mathcal{J}_z=r\,\mathcal{P}_\phi$. However, it is known that the group velocity, and hence the integral Poynting-Abraham momentum, of planar surface plasmon-polaritons tends to zero in the large-frequency limit \cite{bliokh2017optical,bliokh2017optical2,maier2007plasmonics,nkoma1974elementary}: $\left\langle \boldsymbol{\mathcal{P}} \right\rangle\rightarrow 0$ for $\omega\rightarrow\infty$. 
[This is caused by opposite directions of the Poynting vector in the vacuum and metal parts of the surface plasmon and is in sharp contrast
to propagating waves in dielectrics.] Therefore, both the integral azimuthal Poynting vector and Poynting-Abraham AM tend to zero: $\langle\mathcal{J}_z\rangle\simeq r_0\langle\mathcal{P}_\varphi\rangle\rightarrow 0$. 
At the same time, the canonical AM of the metallic-wire modes does not vanish and is well-defined, because all the field components $H_{\phi} =(i/ \sqrt{2})(H^+ e^{i\phi} - H^- e^{-i\phi})$, $E_{r} =(1/ \sqrt{2})(E^+ e^{i\phi} + E^- e^{-i\phi})$, and $E_z$ possess the common phase factor $\text{exp}(i\ell\phi)$ [see Eqs.~(\ref{eq:eigenmode}) and (\ref{eq:eigenmode_out})], subject to the action of the AM operator $-i\partial/\partial\phi$.

Analytical calculations for metallic-wire modes do not produce simple equations similar to Eqs.~(\ref{eq:dielectric_simp}) and (\ref{eq:dielectric_param}) because of the dispersion of the metal and the difference between $\varepsilon$ and $\tilde{\varepsilon}$. However, the geometrical-optics picture of surface plasmon-polaritons propagating along helical rays on the metal-dielectric interface, Fig.~\ref{fig:fig0}(b), allows a simple analytical description of the higher-order mode properties in the paraxial approximation, $k_0r_0 \gg 1$.

Consider a locally-planar surface plasmon-polariton propagating with the wavevector $\mathbf{k}_p = k_z \overline{\mathbf{z}} + k_\phi\overline{\boldsymbol{\varphi}}$, where the local Cartesian coordinates of the interface are attached to the global cylindrical coordinates (the overbars denote the corresponding unit vectors), and $|{\bf k}_p|=k_p$ is the wavenumber of the planar surface plasmon-polariton \cite{bliokh2017optical,bliokh2017optical2,maier2007plasmonics,nkoma1974elementary}. Then, the phase-matching (quantization) condition along the cyclical azimuthal coordinate on the cylindrical surface yields $k_\varphi r_0 = \ell$ \cite{catrysse2009understanding}. In turn, the longitudinal wavevector component determines the propagation constant: $k_z = \beta$. From these relations and known properties of surface plasmon-polaritons \cite{bliokh2017optical,bliokh2017optical2,maier2007plasmonics,nkoma1974elementary}, we derive the dispersion relation for metallic-wire modes:
\begin{equation}
\label{eq:an_metal_modes}
\beta(\omega)\simeq\sqrt{k_p^2(\omega)-\frac{\ell^2}{r_0^2}}, \quad k_p(\omega) = \sqrt{\frac{\varepsilon_1(\omega)}{1+\varepsilon_1(\omega)}}\,\frac{\omega}{c}.
\end{equation}
Remarkably, this is a simple non-transcendental relation without any special functions. The mode group velocity can also be derived either by differentiating Eq.~(\ref{eq:an_metal_modes}) or by taking the  $z$-projection of the group velocity of planar surface plasmons \cite{bliokh2017optical,bliokh2017optical2,maier2007plasmonics,nkoma1974elementary}:
\begin{equation}\label{eq:metal_gv}
v_g\simeq c\frac{(1+\varepsilon_1)^2}{1+\varepsilon_1^2}\sqrt{\frac{\varepsilon_1}{1+\varepsilon_1}-\frac{\ell^2}{k_0^2r_0^2}}.
\end{equation}
The comparison of Eqs.~(\ref{eq:an_metal_modes}) and (\ref{eq:metal_gv}) with the results of exact calculations is shown in Fig.~\ref{fig:fig1}(b). These agree well for $k_0r_0 \gg 1$.

Next, it is known now that planar surface plasmon-polaritons carry \emph{transverse spin AM}, orthogonal to their wavevector $\mathbf{k}_p$ and to the normal to the interface (the $r$-direction in our case) \cite{bliokh2015transverse,bliokh2015spin,bliokh2017optical,bliokh2017optical2, bliokh2012transverse,aiello2015transverse}. 
Therefore, this transverse spin has both a $\phi$-component and a $z$-component. Using the transverse spin calculated for planar surface plasmons in \cite{bliokh2017optical,bliokh2017optical2} and projecting it onto the  $z$-axis, we obtain the following longitudinal spin AM of the metallic-wire mode:
\begin{equation}\label{eq:long_spin}
\frac{\omega\left\langle S_z \right\rangle}{\left\langle W \right\rangle}\simeq \frac{\sqrt{-1-\varepsilon_1}(2+\varepsilon_1)}{1+\varepsilon_1^2}\frac{\ell}{k_0r_0}.
\end{equation}
This equation agrees well with the exact calculations, as shown in Fig.~\ref{fig:fig3}, when $k_0r_0 \gg 1$.

Thus, the geometrical-optics ray picture, supplied with the known properties of planar surface plasmon-polaritons, provides an efficient analytical description for the dispersion and AM properties of the higher-order metallic-wire modes.
\textcolor{black}{Note that our model is based on the simple {\it scalar} quantization condition $k_\varphi r_0 = \ell$ \cite{catrysse2009understanding}. Due to the vector nature of surface plasmon-polaritons, one can further improve it by taking into account the geometric-phase correction \cite{bliokh2010angular}.}

%%%%%%%%%%%%%%%%%%%%%%%%%%%%%%%%%%%%%%%%%%
\section{Conclusion}
%%%%%%%%%%%%%%%%%%%%%%%%%%%%%%%%%%%%%%%%%%
%
We have provided the first self-consistent calculations, both analytical and numerical, of the canonical dynamical properties --- spin/orbital/total angular momenta (AM), momentum, and helicity --- of the eigenmodes of cylindrical waveguides: dielectric fibers and metallic wires. These properties are of major importance for optical communications and information transfer, including AM-based multiplexing \cite{gibson2004free,wang2012terabit,tamburini2012encoding,bozinovic2013terabit,willner2015optical}. Surprisingly, despite the long history of the theoretical and experimental studies of optical waveguides \cite{snyder2012optical,marcuse1972light,pfeiffer1974surface,novotny1994light}, there was no proper description of the AM of the cylindrical guided modes. This is because of the lack, until very recently \cite{bliokh2017optical,bliokh2017optical2}, of consistent theoretical definitions of these quantities (well-studied in free space) in inhomogeneous and dispersive media. Our work fills this important gap.

In particular, we have found the fundamental {\it quantization of the total AM} of eigenmodes of cylindrical waveguides. Although this result looks very natural from the symmetry viewpoint, it has never been obtained explicitly, apart from numerical calculations \cite{le2006angular} for a single fundamental mode in a nondispersive dielectric fiber. Notably, the traditional approach based on the kinetic Poynting (i.e., Abraham) momentum and AM results in very different non-integer AM values, counterintuitive for cylindrically-symmetric systems. Furthermore, the Poynting-Abraham AM {\it vanishes} in the paraxial approximation for metallic-wire modes. This is in strong contrast with the vortex nature of higher-order metallic-wire modes.

We have also calculated the spin and orbital AM of the guided modes. These are noninteger in the general nonparaxial case, because of the spin-orbit interactions induced by the inhomogeneous medium \cite{dooghin1992optical,liberman1992spin,bliokh2004modified,alexeyev2007fiber, golowich2014asymptotic,bliokh2015spin}, but tend to integer values (\ref{eq:parax_dielec}) and (\ref{eq:parax_metal}) in the paraxial regime. Remarkably, we have shown that the spin, orbital, and total AM values are intimately related to the generalized \emph{geometric and dynamical phases} in the mode fields. The laconic relations (\ref{eq:phase_bar}) generalize previous free-space results \cite{van1994commutation,bliokh2010angular,bliokh2014conservation} to the case of inhomogeneous and dispersive optical media.
We have also provided the simplified geometrical-optics model of metallic-wire modes. This model yields approximate analytical expressions for the mode parameters and shows that the spin AM of metallic-wire modes originates from the {\it transverse spin} of surface plasmon-polaritons \cite{bliokh2015transverse,bliokh2017optical,bliokh2017optical2,bliokh2012transverse, aiello2015transverse} propagating along helical trajectories.

Thus, our approach allows one to quantify the most fundamental dynamical properties of the cylindrical modes in the exact full-vector formalism. In all cases we examined, the results are perfectly consistent with the physical intuition and symmetries of the system, see Eqs.~(\ref{eq:group_vel}), (\ref{eq:integer}), (\ref{eq:parax_dielec}), and (\ref{eq:parax_metal}). Therefore, our consideration of cylindrical media can be regarded as a simple \emph{test case} for further application of the general formalism of Eqs.~(\ref{eq:definitions}) and (\ref{eq:helicity}) to optical eigenmodes of complex dielectric and metallic structures. 

\textcolor{black}{After this work was completed, the relevant recent paper \cite{Kien2017} and the preprint \cite{partanen2018angular} came to our attention. The paper \cite{Kien2017} examines the spin and orbital AM, as well as the helicity, of the eigenmodes of nondispersive dielectric fibers. However, the Poynting-Abraham-type quantities are analyzed there, which differ considerably from the canonical Minkowski-type quantities considered in our work. In turn, the preprint \cite{partanen2018angular} reports related results on the quantization of the Minkowski-type total AM of optical beams, but only in homogeneous nondispersive media.}

\subsection*{Funding}
European Research Council (Marie Curie fellowship No. 748950 and Starting Grant ERC-2016-STG-714151-PSINFONI); Air Force Office of Scientific Research (AFOSR) (FA9550-14-1-0040); Japan Science and Technology Agency (JST) (the ImPACT program and CREST Grant No. JPMJCR1676); Japan Society for the Promotion of Science (JSPS) (JSPS-RFBR Grant No. 17-52-50023); John Templeton Foundation; the RIKEN-AIST Challenge Research Fund; the Australian Research Council.

% Bibliography
\bibliography{references}
\bibliographyfullrefs{references}

\clearpage

\section*{Appendix A}
\label{sec:appendix}
\setcounter{equation}{0}
\renewcommand{\theequation}{A\thesubsection\arabic{equation}}

The electromagnetic boundary conditions for $r=r_0$, i.e. the continuity of the $E_{z,\phi}$ and $H_{z,\phi}$ components of the fields (\ref{eq:eigenmode}) and (\ref{eq:eigenmode_out}), provide a system of equations for the coefficients $A$, $B$, $C$, $D$. It can be written as the matrix equation $\hat{M}\,\vec{V}=0$ \cite{marcuse1972light}, with $\vec{V}=\left(A,B,C,D\right)^T$ and:
\begin{eqnarray}
\label{eq:trans_det}
\hat{M}\!=\!\left(\begin{matrix}
\sqrt{\varepsilon_2}J_{\ell} && \hspace{-4mm} 0 && \hspace{-4mm}-\sqrt{\varepsilon_1}H_{\ell}^{(1)} && \hspace{-4mm}0\\
\sqrt{\varepsilon_2}\frac{\ell \beta}{\kappa_1^2 r_0}J_{\ell} &&\hspace{-4mm} i\sqrt{\varepsilon_2}\frac{k_1}{\kappa_1} J_{\ell}' &&\hspace{-4mm} -\sqrt{\varepsilon_1}\frac{\ell \beta}{\kappa_2^2 r_0} H_{\ell}^{(1)} &&\hspace{-4mm} -i\sqrt{\varepsilon_1}\frac{k_2}{\kappa_2} H_{\ell}^{(1)'}\\
0 &&\hspace{-4mm} J_{\ell} &&\hspace{-4mm} 0 &&\hspace{-4mm} -H_{\ell}^{(1)}\\
-i\frac{k_1}{\kappa_1}J'_{\ell} &&\hspace{-4mm} \frac{\ell \beta}{\kappa_1^2 r_0}J_{\ell} &&\hspace{-4mm} i\frac{k_2}{\kappa_2}H_{\ell}^{(1)'} &&\hspace{-4mm} -\frac{\ell\beta}{\kappa_2^2 r_0}H_l^{(1)}
\end{matrix}\right).
\end{eqnarray}
Here, $\kappa_{1,2}=\sqrt{k_{1,2}^2-\beta^2}$, $k_{1,2}=\varepsilon_{1,2}\mu\omega^2/c^2$, $J_{\ell}\equiv J_{\ell}(\kappa_1r_0)$, $H_{\ell}^{(1)}\equiv H_{\ell}^{(1)}(\kappa_2r_0)$, and the prime stands for the derivative with respect to the special-function argument.

The transcendental dispersion equation for the eigenmodes is provided by ${\rm det}\,\hat{M}(\beta,\omega)=0$. After it is solved (numerically), one can find the complex field amplitudes $A$,$B$,$C$, and $D$, up to a common constant factor. In the special case $\ell=0$, Eq.~(\ref{eq:trans_det}) is simplified, and the characteristic equation ${\rm det}\,\hat{M}(\beta,\omega)=0$ can be presented as a product of two factors, one of which must vanish:
\begin{equation}\label{eq:TE_TM}
\frac{J_1}{J_0}-\frac{\varepsilon_2}{\varepsilon_1}\frac{\kappa_1}{\kappa_2}\frac{H_1^{(1)}}{H_0^{(1)}}=0\quad \text{(TM)}, \quad \frac{J_1}{J_0}-\frac{\kappa_1}{\kappa_2}\frac{H_1^{(1)}}{H_0^{(1)}}=0 \quad \text{(TE)},
\end{equation}
where we used $J'_0 = -J_1$, $H_0^{(1)'}=-H_1^{(1)}$, and $k_2/k_1 = \sqrt{\varepsilon_2/\varepsilon_1}$. One can show that these dispersion relations correspond to pure TM and TE modes with $B=D=0$ and $A=C=0$, respectively \cite{marcuse1972light}, and only TM modes exist in the metallic-wire case. Spin, orbital, and total AM, as well as the helicity of the modes \ref{eq:TE_TM}, vanish identically, Eq.~(\ref{eq:ell0}). In the case of dielectric fibers, none of these modes is the fundamental mode with the lowest frequency. The fundamental mode is the circularly-polarized mode characterized by $(m,\sigma,n)=(0,1,0)$, i.e., $\ell =1$, Fig.~\ref{fig:fig1}(a). All modes with $\ell \neq 0$  are generally \emph{mixed}, i.e., neither TE nor TM, with all nonzero coefficients $A$, $B$, $C$, and $D$.

\end{document}